\documentclass[twocolumn,showpacs,preprintnumbers,superscriptaddress]{revtex4}
\usepackage{amssymb}
\usepackage{epsf}

\begin{document}
\draft
\title{Microscopic Derivation of 
Causal Diffusion Equation using Projection Operator Method}
\author{T.~Koide}
\address{Instituto de Fisica, Universidade de Sao Paulo, C.P. 66318, 05315-970 Sao Paulo-SP, Brazil}

\begin{abstract}
We derive a coarse-grained equation of motion of a number density 
by applying the projection operator method to a non-relativistic model. 
The derived equation is an integrodifferential equation 
and contains the memory effect.
The equation is consistent with causality and the sum rule associated with the number conservation 
in the low momentum limit, 
in contrast to usual acausal diffusion equations given by using the Fick's law.
After employing the Markov approximation, 
we find that 
the equation has the similar form to the causal diffusion equation. 
Our result suggests that current-current correlations are not necessarily adequate as 
the definition of diffusion constants.
\end{abstract}


\maketitle

\section{Introduction}

Diffusion is a typical relaxation process and appears in various fields of physics:
thermal diffusion processes, spin diffusion processes, Brownian motions and so on.
It is empirically known that 
the dynamics of these processes is approximately given by the diffusion equation, 
\begin{eqnarray}
\frac{\partial}{\partial t} u({\bf x},t) 
 - D \nabla^2 u({\bf x},t) = 0,
\end{eqnarray}
where $D$ is the diffusion constant.
This equation is phenomenologically derived 
by employing the Fick's law or the Fourier's law.

Although the diffusion equation has broad applicability, 
there exist the limits of the validity.
First of all, the diffusion equation does not obey causality
\cite{ref:Cattaneo,ref:M,ref:IS,ref:IS2,ref:LMR,ref:GL,ref:KL,ref:Jou,ref:Jou2,ref:Az,ref:Az2,ref:Az3,ref:Moha,ref:Kath},
the propagation speed of information exceeds the speed of light.
This means that we cannot apply the ordinary diffusion equation to describe relativistic 
diffusion processes that might be realized in relativistic heavy-ion collisions 
\cite{ref:Az,ref:Az2,ref:Az3,ref:Moha}.
Second,
the diffusion equation breaks sum rules associated with conservation laws\cite{ref:KM}.
The diffusion equation is the coarse-grained equation that is valid only for 
describing macroscopic motions, and hence 
one may claim that such a coarse-grained dynamics does not necessarily satisfy the sum rules.
However, as we will see later, it is possible 
to derive a coarse-grained equation consistent with a sum rule.

These deficiencies can be overcome by introducing relaxation times
\cite{ref:Cattaneo,ref:M,ref:IS,ref:IS2,ref:LMR,ref:GL,ref:KL,ref:Jou,ref:Jou2,ref:Az,ref:Az2,ref:Az3,ref:Moha,ref:KM,ref:Maxwell,ref:MO,ref:Grad,ref:ft1}.
Then, the diffusion equation is changed into the following telegraph equation:
\begin{eqnarray}
\tau \frac{\partial^2}{\partial t^2}u({\bf x},t)
 + \frac{\partial}{\partial t} u({\bf x},t) 
 - D \nabla^2 u({\bf x},t) = 0, \label{eqn:OCDE}
\end{eqnarray}
where $D$ is the diffusion constant and $\tau$ is the relaxation time.
The telegraph equation is reduced into the diffusion equation in the limit of $\tau \longrightarrow 0$.
The propagation speed of the equation is defined by $V=\sqrt{D/\tau}$.
One can easily see that the propagation speed of the diffusion equation diverges 
and hence causality is broken.
Thus, in the following, we call the diffusion equation given by the Fick's law the "acausal" diffusion equation, 
and call the telegraph equation the "causal" diffusion equation.
Furthermore, as we will see later, the causal diffusion equation does not break sum rules.

As just described, the causal diffusion equation may be more appropriate to 
describe diffusion processes \cite{ref:ft2}.
However, the microscopic derivation of the causal diffusion is still 
controversial.
One of the typical methods to derive coarse-grained equations is the projection operator method (POM) 
\cite{ref:Kubo,ref:Fick,ref:KM2,ref:KM3,ref:Naka,ref:Zwanzig1,ref:Mori,ref:SH1,ref:SH2,ref:SH3,ref:SH4,ref:KMT1,ref:KM1,ref:Koide,ref:review,ref:review2}.
In this method, the motions associated with microscopic time and length scales are projected out by 
introducing projection operators, and we can obtain coarse-grained equations expressed in terms of 
variables associated with macroscopic time and length scales.
As is discussed in Ref. \cite{ref:Fick}, 
the coarse-grained equation derived by using the POM 
is not causal diffusion equations but acausal diffusion equations.
However, it should be noted that an approximation whose validity is not obvious is introduced to 
obtain the acausal diffusion equation.
Thus, when we do not use the approximation, 
it may be possible to obtain causal diffusion equations instead of 
acausal ones by using the POM.

As a matter of fact, the author recently applied the POM to derive the coarse-grained 
equation of the order parameter that describes the critical dynamics 
of the chiral phase transition \cite{ref:KM2,ref:KM3}.
The derived equation fulfills the requirements near the critical temperature 
and converges to the equilibrium state consistent with mean-field results 
evaluated in finite temperature field theory.
However, the equation shows the relaxation exhibiting oscillation.
This behavior is different from that of 
the time-dependent Ginzburg-Landau (TDGL) equation 
that has been assumed as a phenomenological equation of the critical dynamics.
We can look upon the TDGL equation as the acausal diffusion equation for "non-conserved" quantities 
because of its overdamping behavior.
Then, the appearance of the oscillation means that a kind of relaxation time is introduced.
This result suggests to us that 
if we apply the POM to derive a coarse-grained equation of a "conserved" quantity, 
for instance, a number density, 
the coarse-grained dynamics may be accompanied by oscillation 
like the causal diffusion dynamics.

In this paper, we apply the POM to a non-relativistic model and show that the coarse-grained equation of 
the number density has a similar form to the causal diffusion equation instead of the acausal one.

This paper is organized as follows.
In Section 2, we summarize the POM.
In Section 3, we apply the POM to a non-relativistic model 
and derive the coarse-grained equation of the number density.
The number density is a conserved quantity in this model 
and there exists the sum rule associated with the conservation law.
We investigate the relation between the coarse-grained equation and the sum rule in Section 4.
The coarse-grained equation contains the memory effect, which can be eliminated by employing the Markov limit.
Then, the equation has a similar form to the causal diffusion equation, as is shown in Section 5.
In the acausal diffusion equation of the number density, 
it is known that the diffusion constant is given by the time correlation 
function of the number density.
However, the simple relation is changed for causal diffusion equations.
The reason is discussed in Section 6.
The summary and concluding remarks are given in Section 7.

\section{Projection operator method}

In this section, we give a short review of the POM 
\cite{ref:Fick,ref:KM2,ref:KM3,ref:Naka,ref:Zwanzig1,ref:Mori,ref:SH1,ref:SH2,ref:SH3,ref:SH4,ref:KMT1,ref:KM1,ref:Koide,ref:review,ref:review2}.
The time-evolution of an arbitrary operator follows the Heisenberg equation of motion, 
\begin{eqnarray}
  \frac{d}{dt}O(t) &=& i[H,O(t)] \\
                   &=& iLO(t) \\
   \longrightarrow O(t) &=& e^{iL(t-t_{0})}O(t_{0}), \label{eqn:HE}
\end{eqnarray}
where $L$ is the Liouville operator and $t_{0}$ is an initial time at which 
we set up an initial state.
To carry out the coarse-grainings of irrelevant information, 
we introduce the projection operators $P$ and its complementary operator 
$Q=1-P$ with the 
following generic properties:
\begin{eqnarray}
   P^2 &=& P, \label{eqn:P} \\
   PQ &=& QP = 0.
\end{eqnarray}
From Eq. (\ref{eqn:HE}), one can see that operators are evolved 
by $e^{iL(t-t_{0})}$, that obeys the following differential equation:
\begin{eqnarray}
   \frac{d}{dt}e^{iL(t-t_{0})} 
   &=& e^{iL(t-t_{0})}iL \nonumber \\
   &=& e^{iL(t-t_{0})}PiL + e^{iL(t-t_{0})}QiL.   \label{eqn:P+Q}
\end{eqnarray}
Operating the projection operator $Q$ from the right, 
we obtain
\begin{eqnarray}
  \frac{d}{dt}e^{iL(t-t_{0})}Q 
&=& e^{iL(t-t_{0})}PiLQ +e^{iL(t-t_{0})}QiLQ. \label{eqn:Q}
\end{eqnarray}
This equation can be solved for $e^{iL(t-t_{0})}Q$,
\begin{eqnarray}
\lefteqn{e^{iL(t-t_{0})}Q } && \nonumber \\
&=& 
Qe^{iLQ(t-t_{0})} 
+ \int^{t}_{t_{0}}d\tau e^{iL(\tau-t_{0})}PiLQe^{iLQ(t-\tau)} 
\label{eqn:TC-dainyuu}.
\end{eqnarray}
Substituting Eq. (\ref{eqn:TC-dainyuu}) 
into the second term on the r.h.s. of Eq. (\ref{eqn:P+Q}) 
and operating $O(t_{0})$ from the right, 
the Heisenberg equation of motion is rewritten as
\begin{eqnarray}
  \frac{d}{dt}O(t) 
&=& e^{iL(t-t_{0})}PiLO(t_{0}) \nonumber \\
&& +\int^{t}_{t_{0}}d\tau e^{iL(t-\tau)}
PiLQe^{iLQ(\tau-t_{0})}iLO(t_{0}) \nonumber \\
&&+ Qe^{iLQ(t-t_{0})}iLO(t_{0}). \label{eqn:TC-1}
\end{eqnarray}
This equation is called the time-convolution (TC) equation.
The first term on the r.h.s. of the equation is called the streaming term 
and corresponds to a collective oscillation such as plasma wave, spin wave, 
and so on.
The second term is the memory term that causes dissipation.
The third term is the noise term.
We can show that 
the memory term can be expressed by the time correlation of the noise.
This relation is called 
the fluctuation-dissipation theorem of the second kind
\cite{ref:Kubo,ref:Fick,ref:Mori,ref:SH1,ref:SH2,ref:SH3,ref:SH4,ref:KM1,ref:Koide,ref:review,ref:review2}.
However, in this paper, 
we simply drop the noise term in the following discussion.

The TC equation (\ref{eqn:TC-1}) is still equivalent to the Heisenberg equation of motion and 
difficult to solve in general.
Note that we can reexpress the memory term of the TC equation as 
\begin{eqnarray}
\int^{t}_{t_{0}}d\tau 
e^{iL(t-\tau)}PiLQ{\mathcal D}(\tau,t_{0})e^{iQL_{0}Q(\tau-t_{0})}iLO(t_{0}),
\end{eqnarray}
where 
\begin{eqnarray}
{\mathcal D}(t,t_{0}) 
&=& 1+\sum^{\infty}_{n=1} i^n \int^{t}_{t_{0}}dt_{1}\int^{t_{1}}_{t_{0}}dt_{2} \cdots \int^{t_{n-1}}_{t_{0}}dt_{n} \nonumber \\
&&\hspace*{-1cm}\times Q\breve{L}^{Q}_{I}(t_{n}-t_{0})Q\breve{L}^{Q}_{I}(t_{n-1}-t_{0}) \cdots 
Q\breve{L}^{Q}_{I}(t_{1}-t_{0}), \nonumber \\
\end{eqnarray}
with
\begin{eqnarray}
\breve{L}^{Q}_{I}(t-t_{0}) 
&\equiv& e^{iQL_{0}Q(t-t_{0})}L_{I}e^{-iQL_{0}Q(t-t_{0})}.
\end{eqnarray}
Here, $L_{0}$ and $L_{I}$ are the Liouville operators of the 
non-perturbative Hamiltonian $H_{0}$ and 
the interaction Hamiltonian $H_{I}$, respectively,
\begin{eqnarray}
L_0~O = [H_0,O],~~~
L_{I}~O = [H_I,O].
\end{eqnarray}
When we expand ${\mathcal D}(t,t_0)$ up to first order in terms of $L_{I}$, 
the TC equation is given by
\begin{eqnarray}
\frac{d}{dt}O(t_{0}) 
&=& e^{iL(t-t_{0})}PiLO(t_{0}) \nonumber \\
&& + \int^{t}_{t_{0}}d\tau e^{iL(t-\tau)}
PiLQe^{iQL_{0}Q(\tau-t_{0})}iLO(t_{0}). \nonumber \\
\label{eqn:TC-2}
\end{eqnarray}
Here, we have already dropped the noise term.
Equation (\ref{eqn:TC-2}) is the starting point in the following calculation.

Note that the memory term contains the coarse-grained time-evolution operator $e^{iQL_{0}Qt}$.
In general, it is not easy to evaluate the coarse-grained time-evolution and hence 
it is approximately replaced by the ordinary time-evolution operator $e^{iL_0 t}$ \cite{ref:Fick},
\begin{eqnarray}
&& \int^{t}_{0}d\tau e^{iL(t-\tau)}
PiLQe^{iQL_{0}Q\tau}iL O(0) \nonumber \\
&& \approx 
\int^{t}_{0}d\tau e^{iL(t-\tau)}
PiLQe^{iL_{0}\tau}iL O(0).\label{eqn:Q-App1}
\end{eqnarray}
See Appendix \ref{App:conv} for details.
This approximation is crucial to derive the "acausal" diffusion equation.
However, as we will see later, it yields several problems.

\section{Application to non-relativistic model}

In this section, we apply the POM to a non-relativistic model.
In the view of causality, coarse-grained equations of non-relativistic models 
is not necessarily causal diffusion equations.
However, if we can implement coarse-grainings preserving conservation laws, 
the causal diffusion equations are more appropriate as coarse-grained equations 
even in the non-relativistic systems, because acausal diffusion equations break sum rules.

We apply the POM to the non-relativistic model with 
the following Hamiltonian:
\begin{eqnarray}
H &=& H_0 + H_I, \label{eqn:Hami}\\
H_0 &=& \int d^3 {\bf x}
\psi^{\dagger}({\bf x}) 
\left(-\frac{1}{2m}\nabla^2-\mu \right) \psi ({\bf x}), \\
H_I &=& -\int d^3 {\bf x} d^3 {\bf x'}
\frac{g}{2}
\psi^{\dagger}({\bf x}) \psi ({\bf x}) v({\bf x-x'}) \psi^{\dagger} ({\bf x'}) 
\psi ({\bf x'}), \nonumber \\
\end{eqnarray}
where $H_0$ and $H_I$ are the non-perturbative Hamiltonian and the interaction 
Hamiltonian, respectively.
The chemical potential $\mu$ is introduced in the non-perturbative Hamiltonian.

The commutation relation of the fermion field $\psi({\bf x})$ is given by 
\begin{eqnarray}
[\psi({\bf x}), \psi^{\dagger}({\bf x'})]_{+}
= \delta^{(3)}({\bf x}-{\bf x'}), \label{eqn:CC}
\end{eqnarray}
where $[~~~]_+$ denotes the anticommutator.

There are many possibilities for the definition of the projection operator 
\cite{ref:Fick,ref:KM2,ref:KM3,ref:Naka,ref:Zwanzig1,ref:Mori,ref:SH1,ref:SH2,ref:SH3,ref:SH4,ref:KMT1,ref:KM1,ref:Koide,ref:review,ref:review2}.
To describe the diffusion process of the number density, 
we employ the following Mori projection operator \cite{ref:Mori},
\begin{eqnarray}
\lefteqn{ P O } && \nonumber \\
&&= \int d^3 {\bf x} d^3 {\bf x}' (O,\delta n({\bf x}))
\cdot (\delta n({\bf x}),\delta n({\bf x}'))^{-1} 
\cdot \delta n({\bf x}'). \nonumber \\
\label{eqn:pro-number}
\end{eqnarray}
Here, $\delta n({\bf x})$ denotes the fluctuations of the number density,
\begin{eqnarray}
\delta n ({\bf x})
= \psi^{\dagger}({\bf x})\psi({\bf x}) 
- \langle \psi^{\dagger} ({\bf x}) \psi({\bf x}) \rangle_{eq}.
\end{eqnarray}
where 
$\langle \psi^{\dagger} ({\bf x}) \psi({\bf x})\rangle_{eq}$ 
is the expectation value 
in thermal equilibrium.
The Kubo's canonical correlation is defined by 
\begin{eqnarray}
(X,Y) = \int^{\beta}_{0}\frac{d\lambda}{\beta}
\langle e^{\lambda H_{0}}Xe^{-\lambda H_{0}}Y \rangle_{0}, \label{eqn:Kubo-C}
\end{eqnarray}
where
\begin{eqnarray}
\langle O \rangle_0 = \frac{1}{Z_0}{\rm Tr}[e^{-\beta H_{0}}O],
\end{eqnarray}
with $Z_0 = {\rm Tr}[e^{-\beta H_0}]$.
When we use this projection operator, we obtain a linear equation.
This is enough for our purpose because we are interested in the dynamics close to equilibrium.
If we want to take nonlinear terms into account, we should re-define a projection operator 
including nonlinear operators. 
See \cite{ref:KM3,ref:Mori} for details.

\begin{figure}\leavevmode
\begin{center}
\epsfxsize=4cm
\epsfbox{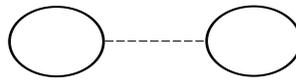}
\caption{
The ring diagram that contributes the calculation of the memory term.
The solid line represents the fermion and the dotted line denotes the interaction $v({\bf x})$.
}
\label{fig:fig1}
\end{center}
\end{figure}

Substituting them into the TC equation (\ref{eqn:TC-2}) and setting $O(0) = \delta n({\bf x})$ and $t_0 = 0$, 
the coarse-grained equation of the number density is derived.
As is discussed in Refs. \cite{ref:KM2,ref:KM3}, the memory function is approximately given by 
calculating the contribution of the ring diagram shown in Fig. \ref{fig:fig1}.
Then, we set $v({\bf x-x'}) = \delta^{(3)}({\bf x-x'})$ for simplicity. 
Finally, we have the integrodifferential equation of the number density,
\begin{eqnarray}
\frac{d}{dt}\delta n({\bf k},t)
=
-
\int^{t}_{0}ds \Gamma({\bf k},t-s) \delta n({\bf k},s). \label{eqn:TC-3}
\end{eqnarray}
Here, the memory function $\Gamma({\bf k},t)$ 
is defined by the inverse-Laplace transform of $\Gamma^L({\bf k},s)$, 
\begin{eqnarray}
\Gamma^L ({\bf k},s)
&=& 
\frac{-\ddot{\chi}^L_s ({\bf k})}{\chi_0 ({\bf k}) + \dot{\chi}^L_s ({\bf k})}
\Omega({\bf k}),  \label{eqn:NRmemory}
\end{eqnarray}
where 
\begin{eqnarray}
\Omega({\bf k})
&=& 1 -g \beta \chi_0 ({\bf k}), \\
\dot{\chi}^{L}_{s}({\bf k}) 
&=& \int^{\infty}_{0} dt e^{-st} 
\frac{d}{dt}\chi_t({\bf k}), \\
\ddot{\chi}^{L}_{s}({\bf k}) 
&=& \int^{\infty}_{0} dt e^{-st} 
\frac{d^2}{dt^2}\chi_t({\bf k}),
\end{eqnarray}
with
\begin{eqnarray}
\lefteqn{\chi_t({\bf k})} && \nonumber \\
&=& 
\frac{1}{\beta V}
\sum_{\bf p}
(n^-_{\bf p+k} - n^-_{\bf p})
\frac{-1}{E_{\bf p+k}-E_{\bf p}}
e^{-i(E_{\bf p+k}-E_{\bf p})t}. \nonumber \\
\end{eqnarray}
Here, $n^-_{\bf k}$ is the Fermi distribution function, 
\begin{eqnarray}
n^-_{\bf k} = \frac{1}{e^{\beta (E_{\bf k} - \mu)}+1},
\end{eqnarray}
with $E_{\bf k} = k^2/2m$.
The function $\chi_t ({\bf k})$ corresponds to the simple contribution of 
the one-loop fermion diagram and is interpreted in the kinematical way as is discussed 
in Refs. \cite{ref:KM2,ref:KM3}.
However, the memory function itself is given by the involved combination of 
$\chi_t({\bf k})$ because of the coarse-grained time-evolution operator.
If we use the approximation discussed in Eq. (\ref{eqn:Q-App1}), 
the memory function is more simplified.
We will come back to this point later.

The Laplace transform of the diffusion equation (\ref{eqn:TC-3}) is 
\begin{eqnarray}
\delta n^L ({\bf k},s)
= \frac{\delta n({\bf k},0)}{s + \Gamma^L({\bf k},s)}.\label{eqn:Laplace-TC3}
\end{eqnarray}
Now, we can investigate whether the number density converges 
to the thermal equilibrium state or not.
When the diffusion equation describes the thermal equilibration process, 
$\delta n({\bf x},t)$ vanishes at late time.
From the final value theorem of the Laplace transformation, 
$\delta n({\bf k},\infty)$ is given by 
\begin{eqnarray}
\lim_{t\rightarrow \infty} \delta n({\bf k},t) 
=\lim_{s \rightarrow 0}  s \delta n^L({\bf k},s).
\end{eqnarray}
Substituting Eq. (\ref{eqn:Laplace-TC3}) into the final value theorem, 
we have
\begin{eqnarray}
\lim_{t\rightarrow \infty} \delta n({\bf k},t) = 0.
\end{eqnarray}
Thus, the time-evolution of $\delta n({\bf k},t)$ is consistent with the fact that 
the derived diffusion equation describes the thermal equilibration process.

As we have discussed at the last paragraph of the preceding section, the projection operator $Q$ 
contained in the memory term is sometimes dropped and the approximation yields several 
problems.
One is the problem of the convergence discussed above.
When we apply the approximation, from Eq. (\ref{eqn:Q-App1}), the memory function is given by 
\begin{eqnarray}
\Gamma^L ({\bf k},s)
= 
- \frac{ \ddot{\chi}^L_s ({\bf k}) }{\chi_0 ({\bf k})} \Omega({\bf k}) . \label{eqn:Q-App2}
\end{eqnarray}
Substituting this expression into the final value theorem,
one can easily check that $\delta n({\bf x},t)$ 
does not vanish at late time,
\begin{eqnarray}
\lim_{t \rightarrow \infty}\delta n({\bf k},t) \neq 0.
\end{eqnarray}
This means that the derived equation cannot describe the 
thermal equilibration process in employing the approximation.
Thus, we should not apply the approximation in this calculation.
Another problem will be discussed in Section 6.

\section{Coarse-grainings and sum rule}

In our Hamiltonian, the number density is a conserved quantity 
and there exists a sum rule associated with the conservation law.
Then, it is desirable that the coarse-grained equation is consistent with 
the sum rule.
In this section, we show that our coarse-grained equation 
is consistent with the sum rule.

First of all, we introduce the Fourier transform of 
the correlation function of the number density \cite{ref:KM},
\begin{eqnarray}
&& \langle [n({\bf x},t), n({\bf x}',t')] \rangle_{eq} \nonumber \\
&& = 
\int\frac{d\omega}{2\pi}
\int\frac{d{\bf k}}{(2\pi)^3}
C''({\bf k},\omega)e^{i{\bf k(x-x')}}e^{-i\omega(t-t')}. \nonumber \\
\end{eqnarray}
The Fourier transform $C''({\bf k},\omega)$ is real and an odd function of the frequency $\omega$ \cite{ref:KM}.
From the sum rule associated with the number conservation, as is discussed in Appendix \ref{App:conser}, 
the Laplace-Fourier transform $\delta n^{LF} ({\bf k},z)$ is expanded for large values of $z$,
\begin{eqnarray}
\lefteqn{ \delta n^{LF} ({\bf k},z)/F({\bf k}) } && \nonumber \\
&=&
\frac{i}{z}C({\bf k}) 
+ \frac{i}{z^3}\frac{1}{m} {\bf k}^2 \langle n ({\bf 0}) \rangle_{eq} + O(1/z^4), 
\label{eqn:expansion}
\end{eqnarray}
where $F({\bf k})$ denotes an external field, and 
\begin{eqnarray}
C({\bf k})
= \int \frac{d\omega'}{2\pi} \frac{C''({\bf k},\omega')}{\omega'}.
\end{eqnarray}
We can see that (i) the term proportional to $1/z^2$ 
disappears and (ii) the coefficient of the term $1/z^3$ is given by 
$i \langle n ({\bf 0}) \rangle_{eq} {\bf k}^2 /m$.
As is shown in Appendix \ref{App:conser}, 
if the dynamics of the number density is approximately given by the acausal diffusion equation, 
one can show that the term proportional to $1/z^2$ does not disappear \cite{ref:KM}.
This is contradiction to the exact result.

On the other hand, the integrodifferential equation (\ref{eqn:TC-3}) obeys 
the sum rule and is consistent with Eq. (\ref{eqn:expansion}).
By setting $s=-iz$ and expanding Eq. (\ref{eqn:Laplace-TC3}) for large values of $z$, 
we obtain 
\begin{eqnarray}
\delta n^{LF} ({\bf k},z)
&=& \frac{i\delta n({\bf k},0)}{z} \nonumber \\
&& + \frac{i\delta n({\bf k},0)}{z^3}
\frac{
-\frac{2}{\beta V}\sum_{\bf p}
n^-_{\bf p}(E_{\bf p+k}-E_{\bf p})
}
{
\frac{1}{\beta V}
\sum_{\bf p}(n^-_{\bf p+k} - n^-_{\bf p})
\frac{1}{(E_{\bf p+k}- E_{\bf p})}
}\Omega({\bf k}) \nonumber \\
&& + O(1/z^4) \nonumber \\
&=&
\frac{i\delta n({\bf k},0)}{z}
+ \frac{i\delta n({\bf k},0)}{z^3}
\frac{
\frac{{\bf k}^2 }{m \beta V}\sum_{\bf p}
n^-_{\bf p}
}
{
\chi_0({\bf k})
}\Omega({\bf k}) \nonumber \\
&& + O(1/z^4). \label{eqn:largezofn}
\end{eqnarray}
One can see that the coefficient of the term proportional to $1/z^2$ vanishes, 
as is the case with Eq. (\ref{eqn:expansion}).

As the initial condition, 
we set 
\begin{eqnarray}
\delta n({\bf k},0)
=
C({\bf k})F({\bf k}),
\end{eqnarray}
where, the function $C({\bf k})$ is given by 
\begin{eqnarray}
\lefteqn{C({\bf k})} && \nonumber \\
&=& \int \frac{d\omega}{2\pi} \frac{\int^{\infty}_{-\infty}  dt \int d^3 {\bf x} 
\langle [\delta n({\bf x},t),\delta n({\bf 0},0)] \rangle_{eq} e^{-i{\bf kx}} e^{i\omega t}}{\omega}. 
\nonumber \\
\label{eqn:CK}
\end{eqnarray}
Substituting it into Eq. (\ref{eqn:largezofn}), we find that 
the first term is identical with that of Eq. (\ref{eqn:expansion}).
On the other hand, we can see the following relation by comparing the second terms:
\begin{eqnarray}
\langle n ({\bf x'}) \rangle_{eq}
= 
C({\bf k})
\frac{
\frac{1}{\beta V}\sum_{\bf p}
n^-_{\bf p}
}
{
\chi_0({\bf k})
}\Omega({\bf k}). \label{eqn:eq-numden} 
\end{eqnarray}
When we apply the free gas approximation to the expectation value of Eq. (\ref{eqn:CK}), 
the initial condition is given by 
\begin{eqnarray}
C({\bf k}) = \beta \chi_0({\bf k}).
\end{eqnarray}
Substituting this expression into Eq. (\ref{eqn:eq-numden}), 
we obtain the number density at the equilibrium state,
\begin{eqnarray}
\langle n ({\bf 0}) \rangle_{eq}
= 
\frac{1}{V}\sum_{\bf p}
n^-_{\bf p}
\Omega({\bf k}).
\end{eqnarray}
The r.h.s. of the equation has the ${\bf k}$ dependence.
It follows that the number density converges to an inhomogeneous distribution instead of 
a thermal equilibrium distribution.
However, it should be noted that the breaking of the sum rule is small for low ${\bf k}$ 
because $\lim_{{\bf k}\rightarrow 0} \Omega({\bf k}) = 1$.
Thus, this approach is still available for 
describing the dynamics associated with long distance scales \cite{ref:ft3}.

\section{Markov approximation}

The derived diffusion equation is still an integrodifferential equation.
To discuss the behavior of the coarse-grained equation, that is, causal or acausal, 
we derive the local equation by employing the Markov approximation.

First of all, we should notice that the memory term can be separated into 
two terms \cite{ref:KM2,ref:KM3},
\begin{eqnarray}
\frac{d}{dt}\delta n({\bf k},t) 
&=& - \int^{t}_{0}d\tau \Omega^2 ({\bf k},t-\tau) \delta n({\bf k},\tau) \nonumber \\
&&- \int^{t}_{0}d\tau \Phi({\bf k},t-\tau) \delta n({\bf k},\tau), 
\end{eqnarray}
where the frequency function $\Omega^2 ({\bf k},t)$ 
and the renormalized memory function $\Phi({\bf k},t)$ 
are defined by the imaginary part and the real part, respectively,
\begin{eqnarray}
\Omega^2 ({\bf k},t)
&=& 
\int \frac{d\omega}{2\pi}i {\rm Im}[\Gamma^L ({\bf k},-i\omega + \epsilon)]
e^{-i\omega t}, \\
\Phi({\bf k},t)
&=& 
\int \frac{d\omega}{2\pi} {\rm Re}[\Gamma^L ({\bf k},-i\omega + \epsilon)]
e^{-i\omega t}.\label{eqn:phi}
\end{eqnarray}
It is worth notifying that the time derivative of the number density 
always vanishes at the initial time, $\left. d(\delta n({\bf k},t))/dt \right|_{t=0}=0$.

From the final value theorem, we can see the temporal behavior of the two functions;
the frequency function converges to a finite value and the renormalized memory function 
vanishes at late time.
Thus, we assume that the renormalized memory function 
relaxes {\it rapidly} and vanishes, while 
the frequency function converges to finite values depending on 
temperatures and chemical potentials after {\it short time evolution}.
Actually, the memory function of the chiral order parameter behaves as is assumed above \cite{ref:KM2,ref:KM3}.
Because we are interested in the slow dynamics associated with the macroscopic time scale, 
we ignore such fast variations.
Then, the frequency function is approximately given by a time-independent constant 
and the time dependence of the renormalized memory function is replaced by 
the Dirac delta function,
\begin{eqnarray}
\Omega^2 ({\bf k},t)
&\approx& \lim_{t\rightarrow \infty}\Omega^2 ({\bf k},t)
\equiv D_{\bf k} {\bf k}^2 , \\
\Phi({\bf k},t)
&\approx& 2 \delta (t) \int^{\infty}_{0}d\tau \Phi({\bf k},\tau)
\equiv \frac{2}{\tau_{\bf k}} \delta(t).
\end{eqnarray}

As a result, the integrodifferential equation is approximately given by
\begin{eqnarray}
\frac{d}{dt}\delta n({\bf k},t)
= 
- D_{\bf k}{\bf k}^2 \int^{t}_{0}d\tau \delta n({\bf k},\tau)
- \frac{1}{\tau_{\bf k}}\delta n({\bf k},t), \label{eqn:lde}
\end{eqnarray}
where
\begin{eqnarray}
D_{\bf k} 
&=& \frac{2\Omega({\bf k})}{m \beta (2\pi)^2 \chi_0 ({\bf k})}
\int^{\Lambda}_{0} d p p^2 n^-_{\bf p}, \label{eqn:dk}\\
\frac{1}{\tau_{\bf k}} 
&=&
|{\bf k}|\chi_0({\bf k})\Omega({\bf k}) \nonumber \\
&&\times \left[ \frac{m \pi}{(2\pi)^2}
\int^{\Lambda}_{0} d p p 
n^-_{\bf p}(1-n^-_{\bf p})\theta(1-\frac{|{\bf k}|}{2p}) \right]^{-1}.
\end{eqnarray}
Here, we introduced the momentum cutoff $\Lambda$.
It should be noted that $D_{\bf k}$ is defined also by the final value theorem,
\begin{eqnarray}
D_{\bf k}{\bf k}^2 = \lim_{s\rightarrow 0} s\Gamma^L({\bf k},s).
\end{eqnarray}
This definition gives completely the same expression as Eq. (\ref{eqn:dk}).

These expressions are further simplified in the low momentum limit,
\begin{eqnarray}
D_{\bf k} &\approx& D \equiv 
\frac{1}{m\beta}\frac{ 
\int^{\Lambda}_{0} d p p^2 n^-_{\bf p} }{\int^{\Lambda}_{0}dp p^2 n^-_{\bf p} (1-n^-_{\bf p})}, 
\label{eqn:D} \\
\frac{1}{\tau_{\bf k}}
&\approx& 
\frac{2}{m\pi}|{\bf k}|\frac{\int^{\Lambda}_{0}dp p^2 n^-_{\bf p} (1-n^-_{\bf p})}{
\int^{\Lambda}_{0} d p p 
n^-_{\bf p}(1-n^-_{\bf p})}.
\end{eqnarray}

For later convenience, we multiply $\tau_{\bf k}$ for both sides.
Finally, the integrodifferential equation is reduced to the following Markovian diffusion equation:
\begin{eqnarray}
\tau_{\bf k} \frac{d^2}{dt^2}\delta n({\bf k},t)
+ \frac{d}{dt}\delta n({\bf k},t)
+ D^{ca}_{\bf k} {\bf k}^2 \delta n({\bf k},t)
=0
,\label{eqn:CDE}
\end{eqnarray}
where 
\begin{eqnarray}
D^{ca}_{\bf k}
\equiv 
D\tau_{\bf k}.
\end{eqnarray}
Solving the differential equation, we should employ the constraint for the initial condition\cite{ref:ft4},
\begin{eqnarray}
\left. \frac{d}{dt}\delta n({\bf k},t) \right|_{t=0} = 0. \label{eqn:const}
\end{eqnarray}
One can easily see that the equation is nothing but the causal diffusion equation.
The diffusion coefficient $D^{ca}_{\bf k}$ corresponds to the diffusion constant of the usual diffusion equation, 
and $\tau_{\bf k}$ represents the relaxation time.
One can easily see that the Markovian diffusion equation is reduced into the 
form of the acausal diffusion equation by setting $\tau_{\bf k} = 0$.

The difference between Eq. (\ref{eqn:OCDE}) and Eq. (\ref{eqn:CDE}) is the 
momentum dependence of the relaxation time.
In Eq. (\ref{eqn:OCDE}), the number density decays 
with the same relaxation time even for the different momentum modes.
On the other hand, our equation shows different relaxation time depending on the momentum.

\section{Diffusion constant and relaxation time}

If the coarse-grained dynamics is given by the acausal diffusion equation, 
the diffusion constant is expressed by the time correlation function of the number density 
(or the time correlation function of the number current density)\cite{ref:Kubo}.
In this section, we give the expressions of the diffusion coefficient and the 
relaxation time in terms of correlation functions.

By using the initial condition given by the linear response of the external field $F({\bf k})$, 
the Laplace-Fourier transform of the causal diffusion equation (\ref{eqn:CDE}) is presented by
\begin{eqnarray}
\delta n^{LF}({\bf k},z)/F({\bf k}) = \frac{(1 - iz\tau_{\bf k})C({\bf k})}{-z^2 \tau_{\bf k} -iz + D_{\bf k} {\bf k}^2}.
\end{eqnarray}
Note that the real part of $\left. \delta n ({\bf k},z) / F({\bf k}) \right|_{z = \omega + i\epsilon}$ 
is related to the correlation function of the 
number density (see Appendix \ref{App:conser}).

Setting $z = \omega + i\epsilon$, we can find that the real part has the following relations;
\begin{eqnarray}
\lim_{\omega \rightarrow 0}\lim_{{\bf k}\rightarrow 0}
\frac{\omega^4}{|{\bf k}|^3}{\rm Re}\frac{n({\bf k},\omega)}{F({\bf k})}
&=& 
\frac{C({\bf 0})D}{\tau}, \\
\lim_{{\bf k} \rightarrow 0}\lim_{\omega \rightarrow 0}
|{\bf k}| {\rm Re}\frac{n({\bf k},\omega)}{F({\bf k})}
&=&
\frac{C({\bf 0})}{D\tau},
\end{eqnarray}
where $\tau = \lim_{{\bf k}\rightarrow 0} (|{\bf k}|\tau_{\bf k})$ and $D$ is defined by Eq. (\ref{eqn:D}).
Thus, it is possible to express the diffusion coefficient and the relaxation time in terms of the 
correlation function.
However, the expressions are not so simple 
as the diffusion constant of the acausal diffusion equation.
This is because our equation is based on the coarse-grained dynamics using the projection operator.
In the POM, the projected microscopic variables are the origin to cause 
the fluctuations and dissipations of variables associated with macroscopic time scales.
Then, the diffusion coefficient is defined by the noise-noise correlation 
instead of the current-current correlation
\cite{ref:Fick,ref:Mori,ref:SH1,ref:SH2,ref:SH3,ref:SH4,ref:KM1,ref:Koide,ref:review,ref:review2}.
As a matter of fact, from Eq. (\ref{eqn:TC-1}), 
the exact TC equation of the number density is given by 
\begin{eqnarray}
\frac{d}{dt}\delta n({\bf x},t) 
&=& \int^{t}_{0}d\tau \int d^3 {\bf x'} 
\Gamma({\bf x-x'},t-s)\delta n({\bf x'},s) \nonumber \\
&& + \xi({\bf x},t).
\end{eqnarray}
Here, the last term represents the noise term that has been ignored until now 
and is defined by 
\begin{eqnarray}
\xi({\bf x},t) = Qe^{iLQt}iL\delta n({\bf x},0).
\end{eqnarray}
The relaxation function $\Gamma({\bf x-x'},t-s)$ is expressed by the noise-noise correlation, 
\begin{eqnarray}
&& \Gamma({\bf x-x'}t-t') \nonumber \\
&& = \int d^3 {\bf x''} (\xi({\bf x},t),\xi({\bf x''},t'))\cdot (\delta n({\bf x''},0),\delta n({\bf x'},0))^{-1}.
\nonumber \\
\end{eqnarray}

In the linear response theory, it is known that the diffusion constant 
is expressed by the correlation function of currents 
when the coarse-grained dynamics is assumed to be given by the acausal diffusion equation\cite{ref:Kubo}.
To derive the acausal diffusion equation after taking the Markov limit in the POM, 
the noise-noise correlation must be replaced by the current-current correlation, 
at least, in the low momentum limit.
This is easily verified if we can approximately replace the coarse-grained time-evolution operator $e^{iLQt}$ 
with the usual time-evolution operator $e^{iLt}$ (see Appendix \ref{App:conv} for details).
As a matter of fact, this approximation means to drop $\dot{\chi}^L_{s}({\bf k})$ 
in the denominator of the memory function, and hence 
the memory function is approximately replaced by Eq. (\ref{eqn:Q-App2}).
Then, one can easily show that the Markovian diffusion equation is reduced to the acausal diffusion equation 
in the low momentum limit.
It is normally assumed that the approximation is justified at least in the low momentum limit \cite{ref:Fick}.
Actually, the famous conclusion that the diffusion constant is given by the current-current correlation 
is derived under the same assumption \cite{ref:reichl}.
However, $\dot{\chi}^L_{s}({\bf k})$ is not small even in the low momentum limit.
This is the reason why we cannot obtain the acausal diffusion equation in contrast to Ref. \cite{ref:Fick}.
This result further suggests that the current-current correlation does not necessarily give 
the definition of the diffusion constant.

\section{Summary and Concluding remarks}

We applied the projection operator method (POM) to the non-relativistic model 
and derived the coarse-grained equation for the number density.
The derived equation is an integrodifferential equation and 
contains the memory effect.
In our model, the number density is a conserved quantity and 
there exists the sum rule associated with it.
The usual acausal diffusion equation breaks the sum rule.
On the other hand, the integrodifferential equation satisfies the sum rule in the low momentum limit.

Second, we assumed that there exists the clear separation between microscopic and macroscopic scales, and 
employed the Markov approximation.
The Markovian diffusion equation is characterized by 
the diffusion coefficient and the relaxation time as is the case with the causal diffusion equation.
Thus, we can conclude that the causal diffusion equation can be derived by using the POM.
However, it should be noted that the relaxation time of our equation depends on the momentum.

To derive the acausal diffusion equation in the POM, as is discussed in Ref. \cite{ref:Fick}, 
we should approximately replace 
the coarse-grained time-evolution operator with the normal time-evolution operator.
Usually, it is assumed that the approximation is justified in the low momentum limit\cite{ref:Fick}.
However, in our calculation, we cannot employ the approximation even in the low momentum limit.
This result further suggests that the current-current correlation does not necessarily give 
the definition of the diffusion constant.

These results are not particular to the model used in this paper.
The causal diffusion equation is obtained also by applying the POM to 
the Nambu-Jona-Lasinio (NJL) model, which is the low energy effective model of quantum chromodynamics (QCD).
This result may be important to discuss the QCD critical dynamics.
Usually, we simply assume acausal diffusion equations as coarse-grained equations of 
conserved variables associated with macroscopic time and length scales 
near the critical points \cite{ref:RW,ref:SS,ref:Fuku,ref:TK}.
However, the coarse-grained equation does not necessarily have an acausal form \cite{ref:KM2,ref:KM3,ref:KKR}.
When we apply causal diffusion equations instead of acausal ones, 
the QCD critical dynamics might be changed.

In this calculations, we employed the perturbative approximation.
Thus, there may exist criticism 
that if we calculate without employing approximations, 
we may be possible to obtain acausal diffusion equations instead of causal ones.
However, it is impossible because the exact calculation must satisfy the sum rule.

Ichiyanagi also discussed the causal diffusion equation 
(more generally, the extended irreversible thermodynamics) 
based on the projection operator method \cite{ref:Ichi1,ref:Ichi2}.
However, the situation discussed by him is different from ours.
He introduced a time-smoothed density matrix to define the projection operator.
The time-smoothed density matrix represents a non-equilibrium state and hence, 
the Kubo's canonical correlation of Eq. (\ref{eqn:Kubo-C}) is replaced by a non-equilibrium expectation value.
In our calculation, we consider the situation where the deviation from the equilibrium state is not so strong and 
the degrees of freedom associated with microscopic time scale has already 
reached the thermal equilibrium.
This is the reason why we employ the Kubo's canonical correlation as the definition of the scalar product 
contained in the projection operator.
On the other hand, 
when we employ the time-smoothed density matrix 
as the definition of the scalar product, 
the microscopic degrees of freedom stays in the non-equilibrium state 
and hence the derived coarse-grained 
equation describes the evolution of the fluctuations from the non-equilibrium state.

Recently, the generalized projection operator method (GPOM) is developed
to describe the pulse parameter dynamics \cite{ref:GPOM}.
Although projection operators are introduced also in the GPOM, 
the GPOM is different from our method.
The Mori projection operator satisfies the condition (\ref{eqn:P}), 
and this plays an important role in deriving the TC equation (\ref{eqn:TC-2}).
However, the projection operator introduced in the GPOM does not satisfy 
this condition because the phase degree of freedom is introduced.
Thus, it is difficult to reproduce the GPOM in our projection operator method.

\begin{acknowledgements}
We acknowledge financial support by FAPESP(04/09794-0).
\end{acknowledgements}

\appendix

\section{Conventional derivation of diffusion equation 
in the projection operator method} \label{App:conv}

In this section, we review the conventional derivation of 
a coarse-grained equation of the number density in the POM \cite{ref:Fick}.

Substituting the Mori projection operator defined by Eq. (\ref{eqn:pro-number}) 
into Eq. (\ref{eqn:TC-1}), the exact TC equation is given by
\begin{eqnarray}
\lefteqn{\frac{d}{dt}\delta n({\bf x},t) } && \nonumber \\
&=&
\int^{t}_{0}d\tau \int d^3 {\bf x}' d^3 {\bf x}''
 (e^{iQL \tau} \delta \dot{n}({\bf x}),\delta \dot{n}({\bf x}')) \nonumber \\
&& \cdot (\delta n({\bf x}'),\delta n({\bf x}''))^{-1} 
 \cdot \delta n({\bf x}'',t-\tau) 
+ \xi({\bf x},t), \nonumber \\
\end{eqnarray}
The streaming term vanishes in this definition of the projection operator.
We assume that the memory function can be approximated as follows:
\begin{eqnarray}
(e^{iQL \tau} \delta \dot{n}({\bf x}),\delta \dot{n}({\bf x}'))
&=& 
(e^{iQL \tau} \nabla_{x}\cdot \delta {\bf J}({\bf x}),
\nabla_{x'}\cdot\delta {\bf J}({\bf x}')) \nonumber \\
&\approx& 
-\nabla_{x}^2 (e^{iL \tau} \delta {\bf J}({\bf x}), \delta {\bf J}({\bf x}')),
\label{eqn:ex-appro}
\end{eqnarray}
where $\delta {\bf J}({\bf x})$ is the current of the number density.
In the first line, 
we used the equation of continuity,
\begin{eqnarray}
\frac{d}{dt}\delta n({\bf x},t) + \nabla\cdot \delta {\bf J}({\bf x},t) = 0.
\end{eqnarray}
In the second line, 
the coarse-grained time-evolution operator $e^{iQL t}$ 
is approximately replaced by 
the usual time-evolution operator $e^{iL t}$.

After the Fourier transformation, the TC equation is given by
\begin{eqnarray}
\frac{d}{dt}\delta n({\bf k},t)
&=&
{\bf k}^2
\int^{t}_{0}d\tau 
\Gamma_{JJ}({\bf k},\tau) \delta n({\bf k},t-\tau) \nonumber \\
&& + \xi({\bf k},t), 
\end{eqnarray}
where $\Gamma_{JJ}({\bf k},t)$ is the Fourier transform of the memory function,
\begin{eqnarray}
&& \int d^3 {\bf x}'
(e^{iL t} \delta {\bf J}({\bf x}),\delta {\bf J}({\bf x}'))
\cdot 
(\delta n({\bf x}'),\delta n({\bf x}''))^{-1} \nonumber \\
&&=
\int\frac{d^3{\bf k}}{(2\pi)^3} \Gamma_{JJ}({\bf k},t)
e^{i{\bf k(x-x')}}. \nonumber \\
\end{eqnarray}

We assume that $\Gamma_{JJ}({\bf 0},t)$ is finite after taking the Markov approximation. 
Finally, 
the coarse-grained equation at low ${\bf k}$ is given 
by 
\begin{eqnarray}
\frac{d}{dt}\delta n({\bf k},t)
= {\bf k}^2 D^{aca} \delta n({\bf k},t) + \xi({\bf k},t),
\end{eqnarray}
where the diffusion constant is defined by
\begin{eqnarray}
D^{aca} = \int^{\infty}_{0}d\tau \Gamma_{JJ}({\bf 0},\tau).
\end{eqnarray}
This is the acausal diffusion equation, and hence it is often claimed that 
acausal diffusion equation can be derived in the POM.
However, as we have seen, 
to derive the acausal diffusion equation, 
we should apply the approximation (\ref{eqn:ex-appro}) and 
it is not applicable to the non-relativistic Hamiltonian used in this paper.

\section{Correlation functions of number density} \label{App:conser}

First of all, we summarize the general properties of the 
correlation functions of the number density following the discussion given by Kadanoff and Martin \cite{ref:KM}.
This discussion is of assistance when we investigate the validity of the 
coarse-grained equation obtained by applying the POM.

In the linear response theory, 
the expectation value of an arbitrary 
operator $\langle O(t) \rangle$ that is dynamically induced 
by the external field is given by
\begin{eqnarray}
\langle O({\bf x},t) \rangle - \langle O({\bf x},t) \rangle_{eq}
= 
i\int^{t}_{-\infty}ds \langle [ H_{ex}(s),O({\bf x},t)] \rangle_{eq},\nonumber \\
\label{eqn:ex-LRT}
\end{eqnarray}
where $\langle~~~\rangle_{eq}$ means to take a thermal expectation value with 
a Hamiltonian $H$, and $O({\bf x},t) \equiv e^{iHt}O({\bf x})e^{-iHt}$.

We are interested in the number density induced by the external field 
$F({\bf x},t)$.
For this purpose, we substitute 
$O({\bf x})= n({\bf x})$ and 
$H_{ex}(t) = -\int d^3 {\bf x} n({\bf x},t) F({\bf x},t)$ into 
Eq. (\ref{eqn:ex-LRT}). 
Then, we obtain 
\begin{eqnarray}
\langle \delta n({\bf x},t) \rangle
= i\int^{t}_{-\infty}dt' \langle [n({\bf x},t),n({\bf x}',t')] \rangle_{eq}
F({\bf x}',t'). \label{eqn:LRND}
\end{eqnarray}
Here, the external field has the following time dependence:
\begin{eqnarray}
F({\bf x},t) = 
\left \{
\begin{array}{cc}
F({\bf x})e^{\epsilon t} & t < 0 \\
0 & t > 0.
\end{array}
\right.
\end{eqnarray}
Now, we introduce a function $C''({\bf k},\omega)$ as follows;
\begin{eqnarray}
&& \langle [n({\bf x},t), n({\bf x}',t')] \rangle_{eq}  \nonumber \\
&& =
\int\frac{d\omega}{2\pi}
\int\frac{d{\bf k}}{(2\pi)^3}
C''({\bf k},\omega)e^{i{\bf k(x-x')}}e^{-i\omega(t-t')}. \nonumber \\
\label{eqn:AbDS}
\end{eqnarray}
The $C''({\bf k},\omega)$ is real and an odd function of the frequency $\omega$.

Substituting Eq. (\ref{eqn:AbDS}) into Eq. (\ref{eqn:LRND}), 
we obtain
\begin{eqnarray}
\lefteqn{ \langle \delta n({\bf x},t) \rangle } && \nonumber \\
& = &
\int \frac{d\omega}{2\pi} \int \frac{d^3 {\bf k}}{(2\pi)^3}
F({\bf k})\frac{C''({\bf k},\omega)}{\omega}e^{i{\bf kx}}e^{-i\omega t}~~~(t \ge 0).\nonumber \\
\label{eqn:LR}
\end{eqnarray}
We further define the Laplace-Fourier transform of the number density, 
\begin{eqnarray}
\delta n^{LF}({\bf k},z) 
&=& \int d^3 {\bf x} \int^{\infty}_{0}dt 
e^{izt}\langle \delta n({\bf k},t) \rangle \nonumber \\
&=& 
\int\frac{d\omega}{2 \pi i}
\frac{C''({\bf k},\omega)}{\omega(\omega - z)}
F({\bf k}).\label{eqn:NLF}
\end{eqnarray}
Here, we substitute Eq. (\ref{eqn:LR}).

The dynamic susceptibility is defined by
\begin{eqnarray}
C({\bf k},z) 
= \int \frac{d \omega'}{2\pi} \frac{C''({\bf k},\omega')}{\omega'-z}.
\label{eqn:DS}
\end{eqnarray}
Setting $z = \omega + i \epsilon$, the dynamic susceptibility 
is decomposed into the real part and the imaginary part;
\begin{eqnarray}
C({\bf k},\omega+ i\epsilon) = C'({\bf k},\omega) + \frac{i}{2}C''({\bf k},\omega), 
\end{eqnarray}
where
\begin{eqnarray}
C'({\bf k},\omega) = P\int \frac{d\omega'}{2\pi}
\frac{C''({\bf k},\omega')}{\omega' - \omega}. \label{eqn:C'}
\end{eqnarray}
This is a Kramers-Kronig relation.
Because the function $C''({\bf k},\omega)$ is an odd function of $\omega$, 
we obtain the following relation from the Kramers-Kronig relation,
\begin{eqnarray}
C({\bf k}) \equiv C({\bf k},\omega = 0)
= \int \frac{d\omega'}{2\pi} \frac{C''({\bf k},\omega')}{\omega'}.
\end{eqnarray}

We can derive a sum rule with the help of the equation of continuity,
\begin{eqnarray}
\frac{\partial }{\partial t} n({\bf x},t) + \nabla\cdot {\bf J}({\bf x},t)=0.
\end{eqnarray}
Operating the time derivative to Eq. (\ref{eqn:AbDS}) and applying the 
equation of continuity, we obtain
\begin{eqnarray}
\frac{\partial}{\partial t}
&& \langle [n({\bf x},t), n({\bf x}',t')] \rangle_{eq} \nonumber \\
&&= 
- \langle [\nabla \cdot {\bf J}({\bf x},t), n({\bf x}',t')] \rangle_{eq} 
\nonumber \\
&&=
-i \int\frac{d\omega}{2\pi}
\int\frac{d{\bf k}}{(2\pi)^3} \omega 
C''({\bf k},\omega)e^{i{\bf k(x-x')}}e^{-i\omega(t-t')}. \nonumber \\
\end{eqnarray}

In the Schr\"odinger field discussed in this paper, 
the current operator ${\bf J}({\bf x},t)$ is given by 
\begin{eqnarray}
{\bf J}({\bf x},t)
=
\frac{i}{2m}(\nabla\psi^{\dagger}({\bf x},t)\cdot \psi({\bf x},t) 
- \psi^{\dagger}({\bf x},t)\nabla\psi({\bf x},t)). \nonumber \\
\end{eqnarray}
Then, one can easily calculate the equal time commutator of the 
current and the number density,
\begin{eqnarray}
\langle [{\bf J}({\bf x},t), n({\bf x}',t)] \rangle_{eq} 
= -\frac{i}{m} \nabla \delta({\bf x-x'})\langle n ({\bf x'}) \rangle_{eq}.
\end{eqnarray}
This equation implies
\begin{eqnarray}
&& - \int\frac{d\omega}{2\pi}
\int\frac{d{\bf k}}{(2\pi)^3} \omega 
C''({\bf k},\omega)e^{i{\bf k(x-x')}} \nonumber \\
&& =
\frac{1}{m} \nabla^2 \delta({\bf x-x'})\langle n ({\bf x' = 0}) \rangle_{eq},
\end{eqnarray}
and hence we can obtain the following f-sum rule,
\begin{eqnarray}
\int\frac{d\omega}{2\pi}
\omega C''({\bf k},\omega) 
=
\frac{1}{m} {\bf k}^2 \langle n ({\bf 0}) \rangle_{eq}. \label{eqn:SR2}
\end{eqnarray}

By expanding Eq. (\ref{eqn:NLF}) for large values of $z$ 
and using the relation and the sum rule, we obtain
\begin{eqnarray}
\delta n^{LF} ({\bf k},z)/F({\bf k})
&=&
\frac{i}{z}C({\bf k}) + \frac{i}{z^2}\int \frac{d\omega}{2\pi} C''({\bf k},\omega) \nonumber \\
&& + \frac{i}{z^3}\int \frac{d\omega}{2\pi} \omega C''({\bf k},\omega)
+ \cdots \nonumber \\
&=&
\frac{i}{z}C({\bf k}) 
+ \frac{i}{z^3}\frac{1}{m} {\bf k}^2 \langle n ({\bf 0}) \rangle_{eq} + O(1/z^4).
\nonumber \\
\label{eqn:Ex-exp}
\end{eqnarray}

\subsection{Acausal diffusion equation}

Here, we assume that the time-evolution of the number density 
follows the acausal diffusion equation.
Then, from the Fick's law, the current is given by 
\begin{eqnarray}
{\bf J}({\bf x},t) = D \nabla n ({\bf x},t),
\end{eqnarray}
where $D$ is the diffusion constant.
Substituting it into the equation of continuity, we obtain the acausal diffusion equation,
\begin{eqnarray}
\frac{\partial}{\partial t} n({\bf x},t) = D \nabla^2 n({\bf x},t).
\end{eqnarray}
After the Laplace-Fourier transformation, the number density is given by 
\begin{eqnarray}
n^{LF}({\bf k},z) = \frac{ n({\bf k},0)}{-iz + Dk^2},
\end{eqnarray}
where, $ n({\bf k},0)$ represents an initial value of the number density.
When the initial value is induced by the external field $F({\bf x},t)$ defined in the preceding section, 
we can set 
\begin{eqnarray}
n({\bf k},0) = C({\bf k})F({\bf k}).
\end{eqnarray}
Expanding the Laplace-Fourier transform for large values of $z$, we have
\begin{eqnarray}
\frac{ n^{LF}({\bf k},z)}{F({\bf k})}
= i\frac{C({\bf k})}{z}
- \frac{C({\bf k})Dk^2}{z^2} + \cdots .
\end{eqnarray}
This expression has the term proportional to $1/z^2$, while 
the corresponding term vanishes in Eq. (\ref{eqn:Ex-exp}).
Furthermore, $C''({\bf k},\omega)$ in the diffusion equation is 
\begin{eqnarray}
C''({\bf k},\omega)
=
\frac{C({\bf k})Dk^2 \omega}{\omega^2 + (Dk^2)^2}.
\end{eqnarray}
This expression fails to satisfy the sum rule (\ref{eqn:SR2}).
Thus, if the coarse-grained dynamics is approximated by the acausal diffusion equation, 
the time-evolution completely breaks the sum rule.

\subsection{Causal diffusion equation}

Kadanoff and Martin pointed out that 
we should introduce a relaxation time 
to satisfy the sum rule\cite{ref:KM}.
Then, the Fick's law is modified as follows;
\begin{eqnarray}
\frac{\partial}{\partial t}{\bf J}({\bf x},t)
= -\frac{1}{\tau}{\bf J}({\bf x},t)-D\nabla n({\bf x},t).
\end{eqnarray}
Substituting this equation into the equation of continuity, 
we obtain
\begin{eqnarray}
\left[
\frac{\partial^2}{\partial t^2} + \frac{1}{\tau}\frac{\partial}{\partial t}
- D\nabla^2 
\right] n({\bf x},t) = 0.
\end{eqnarray}
After using the initial conditions
\begin{eqnarray}
&& n({\bf k},0) = C({\bf k}) F({\bf k}), \\
&& \left. \frac{\partial}{\partial t}n({\bf x},t) \right|_{t=0} = 0,
\end{eqnarray}
the Laplace-Fourier transform of the number density is given by
\begin{eqnarray}
\frac{n^{LF}({\bf k},z)}{F({\bf k})}
&=& \frac{C({\bf k})(1-iz\tau)}{-iz + D{\bf k}^2 - \tau z^2} \nonumber \\
&=& i\frac{C({\bf k})}{z} + i\frac{D{\bf k}^2 C({\bf k})}{\tau z^3} + \cdots.
\end{eqnarray}
In this expression, we do not have the term proportional to $1/z^2$ 
and the equilibrium value of the number density is given by 
$\langle n({\bf 0}) \rangle_{eq} = mC({\bf k})D/\tau$.
The equilibrium number density should not depend on the momentum and hence 
the expression of $\langle n({\bf 0}) \rangle_{eq}$ looks inconsistent.
However, the diffusion equation is the coarse-grained equation and it is valid only 
in the low momentum limit.
Then, the function $C({\bf k})$ is approximately given by $C({\bf 0})$.
Thus, one can conclude that the causal diffusion equation is consistent with the sum rule 
in the low momentum limit.


\begin{thebibliography}{99}
%
\bibitem{ref:Cattaneo}
C.~Cattaneo, Atti.~Semin.~Mat.~Fis.~Univ.~Modena {\bf 3}, 83 (1948).
%
\bibitem{ref:M}
I.~M\"uller, Z.~Phys. {\bf 198}, 329 (1967).
%
\bibitem{ref:IS}
W.~Israel, Ann.~Phys. (N.Y.) {\bf 100}, 310 (1976).
%
\bibitem{ref:IS2}
W.~Israel and J.~M.~Stewart, 
Ann.~Phys.~(N.Y.) {\bf 118}, 341 (1979).
%
\bibitem{ref:LMR}
I.~S.~Liu, I.~M\"uller, and T.~Ruggeri, 
Ann.~Phys.~(N.Y.) {\bf 169}, 191 (1986).
%
\bibitem{ref:GL}
R.~Geroch and L.~Lindblom, 
Ann.~Phys.~(N.Y.) {\bf 207}, 394 (1991).
%
\bibitem{ref:KL}
P.~Kost\"adt and M.~Liu, 
Phys.~Rev. D {\bf 62}, 023003 (2000).
%
\bibitem{ref:Jou}
D.~Jou, J.~Casas-V\'azquez, and G.~Lebon, 
Rep.~Prog.~Phys.~{\bf 51}, 1105 (1988).
%
\bibitem{ref:Jou2}
D.~Jou, J.~Casas-V\'azquez, and G.~Lebon, 
Rep.~Prog.~Phys.~{\bf 62}, 1035 (1999).
%
\bibitem{ref:Az}
A.~Muronga, Phys.~Rev.~Lett. {\bf 88}, 062302 (2002);
{\bf 89}, 159901(E) (2002).
%
\bibitem{ref:Az2}
A.~Muronga, Phys.~Rev.~C {\bf 69}, 034903 (2004).
%
\bibitem{ref:Az3}
A.~Muronga and D.~H.~Rischke, nucl-th/0407114.
%
\bibitem{ref:Moha}
M.~Abdel-Aziz and S.~Gavin, Phys. Rev. C {\bf 70}, 034905 (2004).
%
\bibitem{ref:Kath}
W.~L.~Kath, Physica D {\bf 12}, 375 (1984).
%
\bibitem{ref:KM}
L.~P.~Kadanoff and P.~C.~Martin,
Ann.~Phys.~{\bf 24}, 419 (1963).
%
\bibitem{ref:Maxwell}
J.~C.~Maxwell, Philos.~Trans.~R.~Soc.~London {\bf 157}, 49 (1867).
%
\bibitem{ref:MO}
S.~Machlup and L.~Onsager, Phys.~Rev. {\bf 91}, 1512, (1953).
%
\bibitem{ref:Grad}
H.~Grad, in {\it Hundbuch der Physik}, edited by S.~Flugge, (Springer, Berlin, 1958) Vol. 12.
%
\bibitem{ref:ft1}
The problem of causality can be solved also by introducing nonlinear terms \cite{ref:Kath}.
%
\bibitem{ref:ft2}
As a matter of fact, the deviation from the acausal diffusion equation 
has been discussed in, for instance, thermal diffusion processes, spin diffusion processes and so on \cite{ref:Kubo}.
%
\bibitem{ref:Kubo}
R.~Kubo, M.~Toda, and N.~Hashitsume, 
{\it Statistical Physics II} (Springer-Verlag, Berlin, 1983).
%
\bibitem{ref:Fick}
E.~Fick and G.~Sauermann, {\it The Quantum Statistics of Dynamic Process} (Springer-Verlag, Berlin, 1983).
%
\bibitem{ref:KM2}
T.~Koide and M.~Maruyama, nucl-th/0308025.
%
\bibitem{ref:KM3}
T.~Koide and M.~Maruyama, Nucl.~Phys.~A {\bf 742}, 95 (2004).
%
\bibitem{ref:Naka}
S.~Nakajima, Prog.~Theor.~Phys.~{\bf 20}, 948 (1958).
%
\bibitem{ref:Zwanzig1}
R.~Zwanzig, J.~Chem.~Phys.~{\bf 33}, 1338 (1960).
%
\bibitem{ref:Mori}
H.~Mori, 
Prog.~Theor.~Phys.~{\bf 33}, 423 (1965).
%
\bibitem{ref:SH1}
N.~Hashitsume, F.~Shibata and M.~Shing\=u, 
J.~Stat.~Phys.~{\bf 17}, 155 (1977).
%
\bibitem{ref:SH2}
F.~Shibata, Y.~Takahashi and N.~Hashitsume, 
J.~Stat.~Phys.~{\bf 17}, 171 (1977).
%
\bibitem{ref:SH3}
F.~Shibata and T.~Arimitsu, J.~Phys.~Soc.~Jpn~{\bf 49}, 891 (1980).
%
\bibitem{ref:SH4}
C.~Uchiyama and F.~Shibata, Phys.~Rev.~E {\bf 60}, 2636 (1999).
%
\bibitem{ref:KMT1}
T.~Koide, M.~Maruyama and F.~Takagi, Prog.~Theor.~Phys.~{\bf 101}, 373 (1999).
%
\bibitem{ref:KM1}
T.~Koide and M.~Maruyama, 
Prog.~Theor.~Phys.~{\bf 104}, 575 (2000).
%
\bibitem{ref:Koide}
T.~Koide, Prog.~Theor.~Phys.~{\bf 107}, 525 (2002).
%
\bibitem{ref:review} 
J.~Rau and B.~M\"uller, Phys.~Rep.~{\bf 272}, 1 (1996).
%
\bibitem{ref:review2}
U.~Balucanai, M. Howard Lee, and V.~Tognetti, Phys.~Rep.~{\bf 373}, 409 (2003).
%
\bibitem{ref:ft3}
As a matter of fact, even for the original causal diffusion equation
\cite{ref:Cattaneo,ref:IS,ref:LMR,ref:GL,ref:KL,ref:Jou,ref:Az,ref:Az2,ref:Az3,ref:Moha}, 
we need to take the low momentum limit to preserve the sum rule \cite{ref:KM}.
See Appendix \ref{App:conser}.
%
\bibitem{ref:ft4}
The constraint derived from Eq. (\ref{eqn:lde}) is 
$\left. d(\delta n({\bf k},t))/dt \right|_{t=0} = -\delta n({\bf k},0)/\tau_{\bf k}$.
However, as is discussed below Eq. (\ref{eqn:phi}), we should 
employ Eq. (\ref{eqn:const}).
This difference is caused by the Markov approximation.
%
\bibitem{ref:reichl}
L.~E.~Reichl, {\it A Modern Course in Statistical Physics}, (University of Texas Press, Austin, TX, 1980).
%
\bibitem{ref:RW}
K.~Rajagopal and F.~Wilczek, Nucl.~Phys. {\bf B399}, 395 (1993).
%
\bibitem{ref:SS}
D.~T.~Son and M.~A.~Stephanov, hep-ph/0401052.
%
\bibitem{ref:Fuku}
K.~Ohnishi, K.~Fukushima and K.~Ohta, nucl-th/0409046.
%
\bibitem{ref:TK}
T.~Koide, hep-ph/0411207.
%
\bibitem{ref:KKR}
T.~Koide, G.~Krein and R.~O.~Ramos, in preparation.
%
\bibitem{ref:Ichi1}
M.~Ichiyanagi, J.~Phys.~Soc.~Jpn, {\bf 59}, 1970 (1990).
%
\bibitem{ref:Ichi2}
M.~Ichiyanagi, Prog.~Theor.~Phys. {\bf 84}, 810 (1990).
%
\bibitem{ref:GPOM}
K.~Nakkeeran and P.~K.~A.~Wai, Opt. Commun. {\bf 244}, 377 (2005).
%
\end{thebibliography}
\end{document}